# Macroscopic randomness for quantum entanglement generation

Byoung S. Ham

Center for Photon Information Processing, School of Electrical Engineering and Computer Science, Gwangju Institute of Science and Technology

123 Chumdangwagi-ro, Buk-gu, Gwangju 61005, South Korea

bham@gist.ac.kr

(submitted on March 4, 2021)

**Abstract:**
Quantum entanglement between two or more bipartite entities is a core concept in quantum information areas limited to microscopic regimes directly governed by Heisenberg's uncertainty principle via quantum superposition, resulting in nondeterministic and probabilistic quantum features. Such quantum features cannot be generated by classical means. Here, a pure classical method of on-demand entangled light-pair generation is presented in a macroscopic regime via basis randomness. This conflicting idea of conventional quantum mechanics invokes a fundamental question about both classicality and quantumness, where superposition is key to its resolution.

**Introduction**

In the interference of coherence optics resulting from Young's double slits or a Mach-Zehnder interferometer (MZI), quantum superposition between coherent lights propagating along different paths plays an essential role in the phenomena of the first-order intensity correlation $g^{(1)}$ [1]. In the coherence picture of such path-dependent interference, the $g^{(1)}$ correlation requires indistinguishability or randomness between two fields as shown by the Fresnel-Arago law [2]. However, the Fresnel-Arago law based on polarization bases has no direct relation with the phase basis of $\varphi \in \{0, \pi\}$ of the interferometer. In other words, the which-way information in an interferometer in terms of polarization basis has nothing to do with the common requirement of coherence on a phase basis [3,4]. Thus, complete randomness or indistinguishability in an MZI needs to be redefined in terms of coherence or a phase basis rather than polarization or measurement basis. In a single bipartite entity such as a single photon or a two-level single atom, randomness represents an indistinguishable property in the basis choice for the system. Such a random property in a bipartite system is accomplished via superposition between the fundamental bases comprising the system [5].

In a coupled system composed of two binary bases, joint randomness can also be accomplished by a tensor product of the bases [5]. In an MZI, the which-way information of a pair of photons or fields should be related to which basis is taken. The basis choice is directly related to the deterministic output port information, regardless of whether the input characteristics are for a single photon or not [6]. Similar to the complementarity theory or uncertainty principle in quantum mechanics regarding conjugate variables, such as position and momentum of a photon, the equivalent relation needs to be defined in an entangled system of a photon pair [7]. In an MZI, the counting photon numbers or energy levels in a Fock state is with respect to the particle nature of photons or atoms and cannot be compatible with the wave nature due to the complementarity relation [8,9]. Thus, the particle nature-based photon number cannot be mixed with the phase-based coherence in an MZI for the general discussion of randomness or indistinguishability.

Quantum entanglement between two individual photons in a coupled system can be naturally accomplished via spontaneous parametric down conversion processes (SPDC) in a $\chi^{(2)}$ nonlinear optical medium as shown in Fig. 1(a). In a two-level atomic system, the pulse area $\Phi$ ($= \Omega T$) of a pump field has a basis of $\Phi \in \{0, \pi\}$ to define the state $|S\rangle$ of the atom-field coupled system, $|S\rangle = (a|g\rangle + be^{i\Phi}|e\rangle)/\sqrt{2}$, where $\Omega$ is the Rabi frequency and T is the pulse duration of $\Omega$ [10]. For the basis choice of $\Phi$, the coupled system is determined whether its final state is in the excited or ground state. In the $\chi^{(2)}$ processes, the randomness of the relative phase between signal and idler photon pairs resulting from the SPDC processes is accomplished via a spontaneous emission decay process, resulting in $\pi/2$ phase difference between them [11,12].

**Results**

*Macroscopic randomness in a coupled MZI*

In a coupled MZI, such randomness can also be accomplished via superposition between phase bases. In a single MZI, specific knowledge of the phase basis $\varphi \in \{0, \pi\}$ determines the output port information, resulting in an interferometric fringe (see Fig. 1(b)), where the split fields α and β have a π/2 phase difference [13]. If the phase basis choice is mixed, then the output direction becomes blurred. Thus, the uncertainty or complementarity relation between the phase basis and output ports is not satisfied in the single MZI. Unlike the single MZI, however, the added $\psi$−based dummy MZI negates the $\varphi$−based MZI determinism, resulting in complete randomness:

$$I_C = I_0(1 + sin\varphi sin\psi)/2, \qquad (1)$$
$$I_D = I_0(1 - sin\varphi sin\psi)/2. \qquad (2)$$

The dummy MZI plays a role of superposition in a phase basis domain, resulting in MZI randomness: $I_C = I_D = I_0/2$. This is the macroscopic randomness for the which-way information in a coupled MZI system of Fig. 1(b).

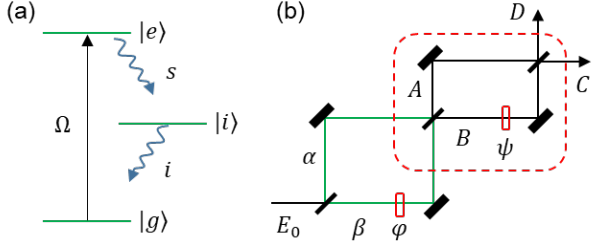

Fig. 1. Schematic of the quantum superposition. (a) a SPDC process. (b) A modified MZI. $\Omega T \in \{0, \pi\}$: Rabi frequency related phase basis. $\varphi, \psi \in \{0, \pi\}$: MZI phase basis.

If we choose $\psi = \pm\pi/2$, Fig. 1(b) turns out to be deterministic again, resulting in a $\varphi$−dependent fringe in both $I_C$ and $I_D$, where the condition of $\psi = \pm\pi/2$ is a direct superposition between the two phase basis of an MZI. Thus, the complementarity relation unsatisfied in a single entity is now satisfied in a coupled MZI. The condition of $\psi = \pm\pi/2$ in Fig. 1(b) results in quantum superposition between entangled pairs, where a single MZI acts as a quantum device as in the SPDC photon pair. Thus, the coupled MZI system of Fig. 1(b) acts as a macroscopically scalable quantum device, satisfying complementarity theory of quantum mechanics: specific knowledge about $\psi$ basis results in complete randomness in the output port decision [3,4]. A similar theory has already been previously proposed [15] and demonstrated in coherence de Broglie wavelength (CBW) [16], which is equivalent to the photonic de Broglie wavelength in a particle nature-based system [17-19]. Unlike CBW, in this manuscript, a novel method of coherence-controlled, on-demand quantum entangled light-pair generation is proposed in a single MZI using phase-basis superposition control, satisfying the complementarity theory in a macroscopic regime.

*Macroscopic entanglement generation in an MZI*

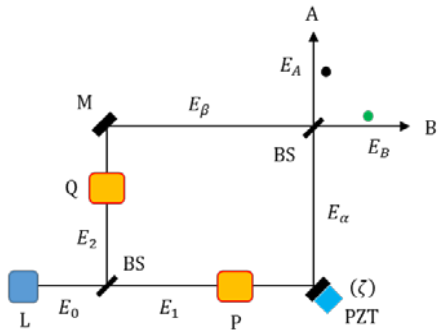

Fig. 2. Schematic of the on-demand entangle light-pair generation. L: laser, BS: beam splitter, M: mirror, PZT: piezo-electric transducer. P and Q: symmetric phase control modules.

Figure 2 shows a schematic of the proposed on-demand coherence control for entangled light-pair generation in a macroscopic regime, where coherence control of the phase bases for the path superposition is key to understanding the fundamental physics of macroscopic quantum features via basis randomness. As mentioned above, the path superposition in an MZI for randomness represents the basis superposition. The phase-basis superposition is accomplished by P and Q in Fig. 2, where the P and Q are phase controllers, resulting in a relative $\delta\varphi$ ($= \varphi' - \varphi''$) phase shift. Thus, the intermediate fields $E_\alpha$ and $E_\beta$ have additional phases of $\varphi'$ and $\varphi''$, respectively. If the controlled phase difference is $\delta\varphi = \pi$ between two fields ($E_\alpha$ and $E_\beta$), the output intensities ($E_A$ and $E_B$) are swapped between A and B ports with respect to the common $\zeta$-controlled case. If the added phases by P and Q are alternatively combined with the uncontrolled input field $E_0$, the output field directions are also alternatively switching between A and B, resulting in an uniform intensity in time average for both output ports. This is equivalent to Fig. 1 in terms of microscopic randomness in a coupled photon system, where such randomness is achieved by superposition between the fundamental phase bases of an MZI.

Figure 3 shows a novel method to satisfy the phase-basis superposition suggested in Fig. 2. In Fig. 3(a), the phase control in each MZI path of Fig. 2 is performed by an acousto-optic modulator (AOM) driven by an rf generator for both P and Q. For rf pulse manipulations, an arbitrary function generator (AFG) is used to adjust the pulse duration T and its duty cycle. The controlled rf pulses are simultaneously supplied to the pair of AOMs to generate $\pm$ first-order deflected optical pulse streams, as shown by the color bar sequences in Fig. 3(a). Both deflected optical pulses at frequencies $f'$ and $f''$, respectively, are combined with the undeflected (zeroth order) optical pulses, resulting in $f' = f_0 + \Delta$ and $f'' = f_0 - \Delta$. The rf pulse duration T is adjusted to be $\Delta T = \pi/2$, resulting in $\varphi' = \pi/2$ and $\varphi'' = -\pi/2$. Thus, phase-basis superposition $\delta\varphi = \pi$ is accomplished. Such phase-basis superposition for $\delta\varphi = \pi$ can also be achieved by an AOM only for Q.

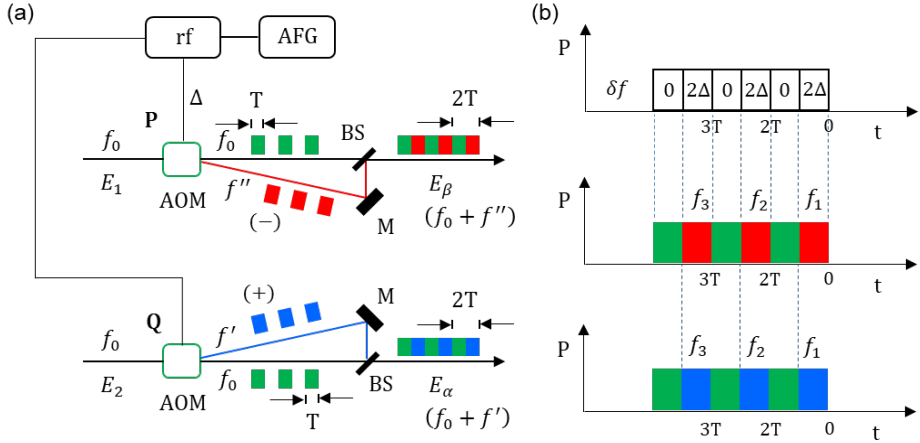

Fig. 3. A method to satisfy the symmetric phase induction in Fig. 1. $f_0$ is the frequency of the input field $E_0$. $\Delta$ is the rf frequency, whose pulse duration (T) and duty cycle are controlled by arbitrary function generator (AFG). The $(+)$ and $(-)$ signs indicate first-order diffraction into positive and negative direction, respectively. $\varphi f = \varphi' - \varphi''$, where $\varphi' = 2\pi\Delta T$ and $\varphi' = -2\pi\Delta T$. AOM: acousto-optic modulator, BS: beam splitter. $f' = f_0 + \Delta$. $f'' = f_0 - \Delta$.

Using matrix representations, the following output fields for Fig. 2 are obtained:
$$\begin{bmatrix} E_A \\ E_B \end{bmatrix} = [BS][Z][\Delta][BS] \begin{bmatrix} E_0 \\ 0 \end{bmatrix}$$
$$= \frac{1}{2}\begin{bmatrix} 1 & i \\ i & 1 \end{bmatrix}\begin{bmatrix} e^{i\zeta} & 0 \\ 0 & 1 \end{bmatrix}\begin{bmatrix} e^{i\varphi'} & 0 \\ 0 & e^{i\varphi''} \end{bmatrix}\begin{bmatrix} 1 & i \\ i & 1 \end{bmatrix}\begin{bmatrix} E_0 \\ 0 \end{bmatrix}$$
$$= \frac{1}{2}\begin{bmatrix} e^{i(\zeta+\varphi')} - e^{i\varphi''} & i[e^{i(\zeta+\varphi')} + e^{i\varphi''}] \\ i[e^{i(\zeta+\varphi')} + e^{i\varphi''}] & -e^{i(\zeta+\varphi')} + e^{i\varphi''} \end{bmatrix}\begin{bmatrix} E_0 \\ 0 \end{bmatrix}. \qquad (3)$$

The corresponding output intensities of MZI in Fig. 2 are as follows:
$$I_A = \frac{I_0}{4}[e^{i(\zeta+\varphi')} - e^{i\varphi''}][e^{-i(\zeta+\varphi')} - e^{-i\varphi''}]$$
$$= \frac{I_0}{2}[1 - \cos(\zeta + \delta\varphi)], \qquad (4)$$

$$I_B = \frac{I_0}{4}\left[e^{i(\zeta+\varphi')} + e^{i\varphi''}\right]\left[e^{-i(\zeta+\varphi')} + e^{-i\varphi''}\right]$$
$$= \frac{I_0}{2}[1 + \cos(\zeta + \delta\varphi)], \tag{5}$$

where $\delta\varphi = \varphi' - \varphi''$. If $\delta\varphi = \pi$ is preset by adjusting $\Delta$ and T for the phase controller P and Q, the output intensities are alternatively swapped with each other for a fixed $\zeta$, whenever the rf pulse is turned on, resulting in $\langle I_B \rangle = \langle I_B \rangle = \frac{I_0}{2}$. Therefore, the randomness condition is satisfied for the MZI, and the resultant intensity correlation $g^{(2)}$ becomes:

$$g^{(2)}(\zeta) = \frac{\langle I_A I_B \rangle}{\langle I_A \rangle \langle I_B \rangle} = \sin(\zeta + \delta\varphi)^2. \tag{6}$$

Unlike alternating output intensities in Equations (4) and (5), the intensity correlation $g^{(2)}(\zeta)$ in Equation (6) is not alternating. Thus, Equation (6) becomes $\zeta$ dependent only, satisfying perfect indistinguishability of the which-way information in an MZI. Notably, the variable $\zeta$ is on the order of wavelength of the input field $E_0$, which is much more sensitive than the conventional results in $\tau-$basd HOM dips [14], where $\tau$ is the bandwidth-caused decay time between paired photons for coincidence measurements. Unlike conventional particle nature-based intensity correlation measurements [7,14], Fig. 2 generates on-demand entangled light pairs in a macroscopic regime, where its operation is deterministic and dynamic. The phase $\zeta$ can be used for system stabilization for the on-demand quantum features via feedback control [20]. The result of macroscopic quantum features in Equation (6) is unprecedented and has potential for future coherence-quantum technologies.

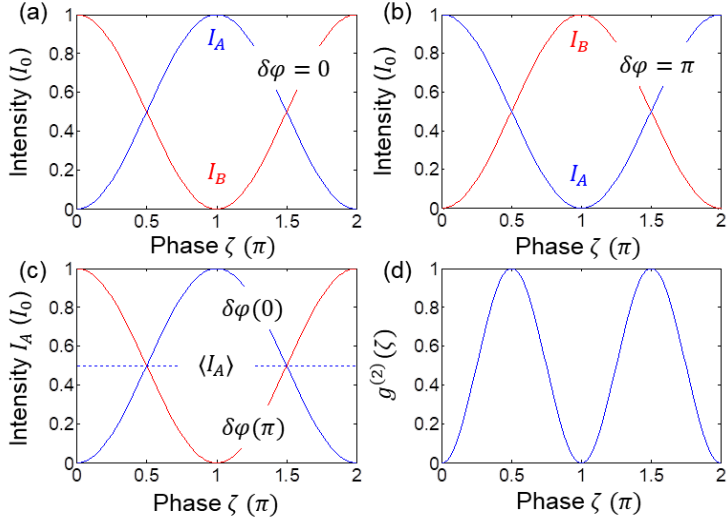

Fig. 4. Numerical calculations for equations (4)-(6).

Figure 4 shows numerical calculations for Equations (4)-(6). Figure 4(a) represents typical MZI output intensities without rf fields, which corresponds to the green pulse set in Fig. 3 for the classical bound. Figure 4(b) is for $\delta\varphi = \pi$, resulting in intensity swapping between $I_A$ and $I_B$, which corresponds to the red and blue pulse set in Fig. 3. Figure 4(c) represents $\delta\varphi-$dependent alternating intensities in the output port A in Fig. 2, resulting in $\langle I_A \rangle = I_0/2$ (see the dotted line). In the output port B, the same alternative and fixed intensity relations are also satisfied. Therefore, according to the definition of intensity correlation, Fig. 4(d) is obtained, where anticorrelation results at $\zeta = n\pi$, regardless of the rf pulse-dependent alternation in each output port. This is equivalent to N=2 case in PBW [18]. All generation processes are classical. The light source used for Fig. 4 is coherent. Thus, the deterministic and macroscopic quantum feature manipulation has been successfully achieved in a single MZI via phase-basis randomness.

**Conclusion**

Unlike conventional probabilistic and microscopic quantum feature generations based on the particle nature of photons, deterministic and macroscopic quantum feature manipulations were proposed, analyzed, and numerically demonstrated for an rf phase-controlled single MZI system. Unlike determinacy between the phase basis and output fields in a typical single MZI, the rf pulse-controlled phase-basis superposition resulted in perfect randomness satisfying the complementarity theory of quantum mechanics in a macroscopic regime. This randomness satisfaction as the wave nature of photons is equivalent to the SPDC generated entangled photon pairs as the particle nature-based one. Potential applications of the proposed method are various from quantum key distribution to quantum sensing using any commercially available laser system, where an extremely high N number in a N00N state is beneficial from critical limitations of photon loss in conventional quantum information.

**References**
1. L. Mandel and E. Wolf, Optical coherence and quantum optics. Ch. 4 (Cambridge university. NY 1995).
2. M. Henry, "Fresnel-Arago laws for interference in polarized light: A demonstration experiment," Am. J. Phys. **49**, 690-691 (1981).
3. D. M. Greenberger, M. A. Horne, and A. Zeiliinger, "Multiparticle interferometry and the superposition principle," Phys. Today **56**(8), 22-29 (1993).
4. S. Dürr, T. Nonn, and G. Rempe, "Origin of quantum-mechanical complementarity probed by a 'which-way' experiment in an atom interferometer," Nature **395**, 33-37 (1998).
5. B. S. Ham, "Analysis of nonclassical features in a coupled macroscopic binary system," New J. Phys. **22**, 123043 (2020).
6. P. Grangier, G. Roger, and A. Aspect, "Experimental evidence for a photon anticorrelation effect on a beam splitter: A new light on single-photon interference," Europhys. Lett. **1**, 173-179 (1986).
7. X. Guo, C.-L. Zou, C. Schuck, H. Jung, R. Cheng, and H. X. Tang, "Parametric down-conversion photon-pair source on a nanophotonic chip," Light: Sci. & Appl. **6**, e16249 (2016).
8. B. T. H. Varcoe, S. Brattke, M. Weidinger, and H. Walther, "Preparing pure photon number states of the radiation field," Nature **403**, 743-746 (2000).
9. M. Hofheinz et al., "Generation of Fock states in a superconducting quantum circuit," Nature **454**, 310-314 (2008).
10. M. O. Scully and M. S. Zubairy, Quantum optics. Ch. 5 (Cambridge, 1999).
11. B. S. Ham, "The origin of anticorrelation for photon bunching on a beam splitter," Sci. Rep. **10**, 7309 (2020).
12. B. S. Ham, "Coherently controlled quantum features in a coupled interferometric scheme," arXiv:2102.09189 (2021).
13. V. Degiorgio, "Phase shift between the transmitted and the reflected optical fields of a semireflecting lossless mirror is π/2," Am. J. Phys. **48**, 81–82 (1980).
14. C. K. Hong, Z. Y. Ou, and L. Mandel, "Measurement of subpicosend time intervals between two photons by interference," Phys. Rev. Lett. **59**, 2044-2046 (1987).
15. B. S. Ham, "Deterministic control of photonic de Broglie waves using coherence optics," Sci. Rep. **10**, 12899 (2020).
16. B. S. Ham, "Observations of coherence de Broglie waves," arXiv:2007.04738 (2020).
17. J. Jacobson, G. Gjork, I. Chung, and Y. Yamamato, "Photonic de Broglie waves," Phys. Rev. Lett. **74**, 4835–4838 (1995).
18. P. Walther *et al.*, "Broglie wavelength of a non-local four-photon state," Nature **429**, 158–161 (2004).
19. X.-L.Wang *et al.*, "18-qubit entanglement with six photons' three degree of freedom," Phys. Rev. Lett. **120**, 260502 (2018).
20. G. B. Xavier and J. P. von der Weid, "Stable single-photon interference in a 1 km fiber-optic Mach-Zehnder interferometer with continuous phase adjustment," Opt. Lett. **36**, 1764-1766 (2011).

Acknowledgments
BSH acknowledges that this work was supported by the "Practical Research and Development support program by GTI (GIST Technology Institute)" funded by GIST in 2021.